\documentclass[sigconf, authorversion]{acmart}
\AtBeginDocument{%
  }

\setcopyright{acmlicensed}

\copyrightyear{2026}
\acmYear{2026}
\setcopyright{cc}
\setcctype{by}
\acmConference[CHI '26]{Proceedings of the 2026 CHI Conference on Human Factors in Computing Systems}{April 13--17, 2026}{Barcelona, Spain}
\acmBooktitle{Proceedings of the 2026 CHI Conference on Human Factors in Computing Systems (CHI '26), April 13--17, 2026, Barcelona, Spain}
\acmPrice{}
\acmDOI{10.1145/3772318.3791517}
\acmISBN{979-8-4007-2278-3/2026/04}

\usepackage{kotex}
\usepackage{framed}
\usepackage{CJKutf8}
\newcommand{\sysname}{\textsc{WriteAid}}

\begin{document}

\title[Investigating Student Interaction with LLM-Based Writing Support in Real-Time K-12 EFL Classrooms]{When Scaffolding Breaks: Investigating Student Interaction with LLM-Based Writing Support in Real-Time K-12 EFL Classrooms}

\author{Junho Myung}
\orcid{0009-0001-8603-1195}
\affiliation{%
  \institution{School of Computing\\KAIST}
  \city{Daejeon}
  \country{Republic of Korea}}
\email{junho00211@kaist.ac.kr}

\author{Hyunseung Lim}
\orcid{0000-0002-5645-1009}
\authornote{Both authors contributed equally to this research.}
\affiliation{
  \institution{Department of Industrial Design\\KAIST}
  \city{Daejeon}
  \country{Republic of Korea}}
\email{charlie9807@kaist.ac.kr}

\author{Hana Oh}
\orcid{0009-0006-2906-0308}
\authornotemark[1]
\affiliation{
  \institution{Department of Intelligence and Information\\Seoul National University}
  \city{Seoul}
  \country{Republic of Korea}}
\email{hana2001@snu.ac.kr}

\author{Hyoungwook Jin}
\orcid{0000-0003-0253-560X}
\affiliation{%
  \institution{Computer Science and Engineering\\University of Michigan}
  \city{Ann Arbor, Michigan}
  \country{United States}}
\email{jinhw@umich.edu}

\author{Nayeon Kang}
\orcid{0000-0002-5781-7359}
\affiliation{%
  \institution{Gyeonggido Office of Education}
  \city{Suwon}
  \country{Republic of Korea}}
\email{nayeonkang@korea.kr}

\author{So-Yeon Ahn}
\orcid{0000-0003-2718-0999}
\affiliation{%
  \institution{School of Digital Humanities and
Computational Social Sciences\\KAIST}
  \city{Daejeon}
  \country{Republic of Korea}}
\email{ahnsoyeon@kaist.ac.kr}

\author{Hwajung Hong}
\orcid{0000-0001-5268-3331}
\affiliation{%
  \institution{Department of Industrial Design\\KAIST}
  \city{Daejeon}
  \country{Republic of Korea}}
\email{hwajung@kaist.ac.kr}

\author{Alice Oh}
\orcid{0000-0002-7884-3038}
\affiliation{%
  \institution{School of Computing\\KAIST}
  \city{Daejeon}
  \country{Republic of Korea}}
\email{alice.oh@kaist.edu}

\author{Juho Kim}
\orcid{0000-0001-6348-4127}
\affiliation{
  \institution{School of Computing\\KAIST}
  \city{Daejeon}
  \country{Republic of Korea}
}
\affiliation{
  \institution{SkillBench}
  \city{Santa Barbara, CA}
  \country{USA}
}
\email{juhokim@kaist.ac.kr}

\renewcommand{\shortauthors}{Myung et al.}

\begin{abstract}

Large language models (LLMs) are promising tools for scaffolding students' English writing skills, but their effectiveness in real-time K-12 classrooms remains underexplored. Addressing this gap, our study examines the benefits and limitations of using LLMs as real-time learning support, considering how classroom constraints, such as diverse proficiency levels and limited time, affect their effectiveness. We conducted a deployment study with 157 eighth-grade students in a South Korean middle school English class over six weeks. Our findings reveal that while scaffolding improved students' ability to compose grammatically correct sentences, this step-by-step approach demotivated lower-proficiency students and increased their system reliance. We also observed challenges to classroom dynamics, where extroverted students often dominated the teacher's attention, and the system's assistance made it difficult for teachers to identify struggling students. Based on these findings, we discuss design guidelines for integrating LLMs into real-time writing classes as inclusive educational tools.

\end{abstract}





\maketitle

\section{Introduction}
\label{sec:intro}

The integration of AI-based learning tools in K-12 education has gained significant attention due to their scalability and potential to enhance learning experiences. These tools provide immediate, adaptive feedback at scale, enabling educators to support a larger number of students while addressing individual learning needs \cite{Gligorea2023Adaptive, Govea2023Optimization}. The need for such tools is particularly evident in EFL classrooms, where students benefit from individualized and immediate feedback on tasks such as essay writing \cite{marzuki2023impact, mohamed2024exploring}. Studies have demonstrated the effectiveness of implementing AI-assisted tools---such as machine translators and Grammarly\footnote{https://www.grammarly.com/}---in classrooms by enhancing students' writing skills through grammar correction, vocabulary suggestions, and improvements in coherence and clarity \cite{crompton2024ai, marzuki2023impact, mohebbi2025enabling}.

The emergence of large language models (LLMs) has created new opportunities in EFL education due to their ability to understand contextual information and provide assistance with sufficient explanations \cite{gan2023large, lee2023learning}. Research has explored how LLMs can offer scaffolding to learners, addressing the limitations of earlier AI-based tools that provided only generic or readily available feedback \cite{yin2024scaffolding, heickal2024generating, liu2024scaffolding}. 

Most existing studies have examined LLM implementation in asynchronous learning environments, where learners interact with the models independently to complete assignments \cite{han2023recipe, wang2024exploring}.
However, the practical realities of in-class, real-time settings introduce a more complicated picture. These environments are shaped by time constraints and dynamic interactions among teachers and peers, where students bring diverse levels of proficiency, motivation, and participation. In such contexts, how we expect LLMs to scaffold learning may backfire; for instance, by overwhelming less proficient learners.
Despite the importance of this tension between the expectations of AI scaffolding and the realities of K-12 classrooms, limited research exists on how LLMs function as real-time support tools and how these unique constraints shape their efficacy.

To this end, we developed \sysname{}, an LLM-based writing support tool that serves as a technology probe (TP) \cite{hutchinson2003technology, huang2014technology} to explore the experiences of K-12 students and their teachers in a real-world classroom setting. Aligned with the teacher's pedagogical requirements and the students' learning context, the system was designed to provide step-by-step guidance on their writing composition through a chat-based interface.

We deployed \sysname{} with eighth-grade students over a six-week period, during which they completed two separate writing tasks. A total of 157 students used the system, and 133 of them consented to share their data for analysis. As a result, we collected 14,863 query-response pairs with a total of 3,733 conversation threads from student-LLM interactions. After fully anonymizing the data, we publicly share the interaction data for research purposes to support future work and further analysis.

To examine the types of queries K-12 students ask LLMs and how they respond to step-by-step guidance, we conducted qualitative coding of interaction data from a randomly selected subset of 15 students with varying levels of English proficiency. Additionally, to explore the role of LLMs in a real-time classroom setting and gain insights beyond interaction data, one of the authors conducted a naturalistic observation study. Three 45-minute writing sessions were observed without intervention, carefully documenting student behaviors and the role of the LLM.

Our qualitative coding of interaction data indicates that while step-by-step guidance from the LLM benefits higher-proficiency students, it may slowly demotivate lower-proficiency students and increase their dependence on the system. Observations from the classroom study show that a few extroverted students dominated interactions with the teacher, while many others turned to the LLM instead of asking questions aloud. This suggests that students who were already less inclined to speak up became even less likely to do so when an approachable alternative was available. Furthermore, as the LLM improves students’ baseline performance, it becomes more challenging for teachers to identify students who are struggling. Based on these findings, we propose design goals and guidelines aimed at ensuring the fair use of LLMs in classroom settings.

The key contributions of this work are as follows:
\begin{itemize}
\item We provide a large-scale dataset\footnote{https://github.com/JunhoMyung/WriteAid-Data} (14,863 utterances) from a real-time K-12 classroom, and reveals a divergence in how students use the LLM: higher-proficiency students delegate lower-order tasks (e.g., vocabulary translation), while lower-proficiency students outsource core higher-order writing tasks (e.g., sentence generation).
\item We identify limitations in LLM scaffolding, revealing that for lower-proficiency students in time-constrained classrooms, it can be counterproductive by fostering dependency and demotivation.
\item Through a naturalistic observation study, we uncover new challenges related to the use of LLMs in educational contexts, particularly concerning issues of fairness and the impact on student-teacher dynamics. 
\end{itemize}

\section{Related Work}
This section reviews prior research on EFL writing in K-12 classrooms, the use of AI tools to support language learning, and the role of LLM-based scaffolding in providing adaptive guidance.

\subsection{Challenges of Teaching Writing in K-12 EFL}
Writing is widely recognized as one of the most cognitively demanding skills in second-language acquisition, integrating grammar, vocabulary, and discourse-level fluency that are essential for academic and communicative competence \cite{wolfersberger2003l1, peter2013tracking}. Yet in K-12 contexts, teachers face persistent structural barriers: large class sizes, limited instructional time, and heterogeneous proficiency levels make it difficult to provide the individualized attention needed to support diverse learners \cite{ghufron2018role, dizon2024systematic}. These challenges limit opportunities for interactive, feedback-rich writing instruction and create a persistent gap between pedagogical ideals and classroom realities\cite{park2019implications, miranty2021automated}.

Pedagogical research highlights three major approaches to EFL writing instruction. Product-oriented approaches emphasize the accuracy of the final essay, aligning with assessment regimes but offering limited support for idea generation or revision \cite{nova2018utilizing, abu_guba2024grammarly}. Process-oriented approaches instead foreground the stages of writing---brainstorming, drafting, feedback, and redrafting---demonstrating clear benefits for learner confidence and writing quality, though they require significant teacher time and attention \cite{almusharraf2023error, hellman2019scaling}. A third approach, genre-based instruction, scaffolds learners through models of target text types (e.g., argumentative essays or reports), offering structured pathways that are effective for EFL learners but equally resource-intensive \cite{ghufron2018role, crompton2024ai}.

In practice, East Asian classrooms often rely on teacher-centered, grammar-focused instruction, leaving little room for peer review, individualized feedback, or formative assessment \cite{marzuki2023impact, mohebbi2025enabling}. Consequently, students frequently complete writing tasks with minimal scaffolding and receive feedback that is delayed or not tailored to their proficiency level \cite{cheng2021_feedback}. This persistent feedback gap has motivated researchers to explore complementary supports—digital platforms, automated writing evaluation systems \cite{ding2024automated}, and more recently AI-based approaches that can deliver just-in-time, adaptive guidance at scale without disrupting whole-class instruction \cite{lv2021onlinefeedback, zhang2023K12Review}.

\subsection{AI Tools for EFL Writing Support}
Digital tools have long been used to support EFL writing, yet they remain limited in their ability to adapt to individual learners' needs. Early systems—such as machine translation \cite{oNeill2019stop}, grammar checkers \cite{nova2018utilizing}, and automated writing evaluation (AWE) platforms \cite{abu_guba2024grammarly, ding2024automated}—provided quick corrections on grammar, spelling, and vocabulary. These tools helped reduce teacher workload and improved surface-level accuracy \cite{karatay2024awesynthesis}, but their feedback was often generic and offered little support for higher-order writing concerns such as idea development, coherence, and argument structure \cite{ghufron2018role, dizon2024systematic}. Moreover, they did not address cognitive challenges specific to L2 learners, such as tip-of-the-tongue states or the mental effort involved in translating from the first language \cite{wolfersberger2003l1, peter2013tracking}. As a result, they functioned primarily as post-hoc correction tools rather than interactive scaffolds for the writing process.

The recent emergence of LLMs has significantly expanded the possibilities for digital writing support \cite{gan2023large, wang2024large}. Unlike earlier rule-based or template-driven systems, LLMs can engage in interactive dialogue and deliver context-aware, adaptive feedback in real time—explaining grammar rules, suggesting vocabulary, and tailoring responses to the learner’s immediate needs \cite{kohnke2023chatgpt, lin2023chatacts, wang2023chatgpt, kasneci2023chatgpt}. Early studies report that LLM-supported writing activities can increase student engagement, facilitate brainstorming, and enhance fluency \cite{han2023recipe, nur2023artificial, sevetlana2024graduate}. At the same time, concerns remain around factual accuracy, academic integrity, and the risk of over-reliance on automated feedback \cite{smolansky2023educator, marzuki2023impact, mohebbi2025enabling}. These challenges highlight the need for careful integration that balances immediate support with opportunities for independent learning \cite{crompton2024ai, li2024languageReview}.

Importantly, most existing research on LLM-assisted writing has focused on adults in higher education or self-study contexts \cite{li2024languageReview, andy2024human, sevetlana2024graduate}. University students, for instance, use LLMs to brainstorm ideas, refine style, and check grammar, but these contexts differ substantially from K–12 classrooms, where lessons are shorter, more structured, and teacher orchestration plays a central role \cite{gao2024a, wangc2024exploring}. Few studies examine the sustained effects of LLM use on grammar or vocabulary retention, and even fewer explore how such tools can be embedded into real-time classroom workflows for middle school learners \cite{kim2024llm}. This gap underscores the need to investigate how LLMs can be designed to function as scalable, pedagogically sound scaffolds that complement teacher feedback and promote long-term writing development in K–12 settings \cite{yin2024scaffolding, malik2024scaling, pal2024autotutor}.

\subsection{LLMs as Scaffolds in the Classroom}
Scaffolding is a well-established concept in education, describing how teachers guide students through tasks that they cannot yet complete independently but can accomplish with structured support \cite{wood1976tutoring, van2010scaffolding}. Building on Vygotsky’s zone of proximal development (ZPD), researchers have identified three key functions of scaffolding: diagnosing the learner’s current level, providing contingent support, and gradually fading assistance as competence grows \cite{khaliliaqdam2014zpd, van2010scaffolding}. In writing instruction, scaffolding strategies often include modeling exemplar texts, providing sentence starters or prompts, offering immediate formative feedback, and gradually removing such supports as learners gain proficiency\cite{taylor2024scaffolding}. Effective scaffolding has been shown to improve learner confidence, retention, and autonomy \cite{damanhouri2021effectiveness}, but in large classrooms teachers often struggle to diagnose each learner’s needs in real time and to deliver individualized, timely support \cite{belland2014scaffolding, haruehansawasin2018scaffolding}.

Recent advances in LLMs create new possibilities for bringing scalable scaffolding into classroom practice. LLMs can provide just-in-time hints, model exemplar solutions, adapt feedback to students’ proficiency levels, and break complex writing tasks into smaller, manageable steps \cite{yin2024scaffolding, malik2024scaling, smith2024prompting}. Such adaptive support has the potential to increase students’ behavioral and cognitive engagement \cite{fincham2024llms, wang2024exploring}, free teachers from repetitive micro-level corrections, and allow them to focus on higher-order writing instruction \cite{han2023recipe, kim2024llm}. Rather than replacing teacher feedback, LLM-based scaffolding can complement it by offering continuous, individualized support between teacher-led interventions \cite{pal2024autotutor, sonkar2024pedagogical}.

Introducing LLM-based scaffolding may also reshape classroom dynamics. As LLMs take on more of the direct support role, teachers’ responsibilities may shift toward orchestrating classroom activities and monitoring students’ use of AI tools \cite{deoliveira2023interactional, kazemitabaar2024codeaid}. This reallocation of teacher effort could enable more time for formative assessment, peer review, and conferencing \cite{mohebbi2025enabling}, while maintaining teacher oversight. However, such changes also raise design considerations around ensuring equitable access, aligning AI feedback with curriculum goals, and maintaining students’ sense of agency \cite{crompton2024ai}.

Despite these possibilities, empirical evidence on LLM-based scaffolding in K–12 EFL classrooms remains scarce \cite{zhang2023K12Review}. We know little about how students at different proficiency levels interact with LLMs or how teacher roles evolve when AI provides step-by-step guidance \cite{li2024languageReview, belland2014scaffolding}.

\section{System Design}


A technology probe (TP) is a tool deployed in real-world settings to explore the unknown and inform new design opportunities \cite{hutchinson2003technology, huang2014technology}. To understand how K-12 students engage with LLM-based writing assistants in a real-time classroom, we developed \sysname{} as a TP.
\sysname{}, an LLM-based scaffolding writing support tool, allows students to freely engage with LLM in a chat-based interface, to get help with sentence construction, grammar revision, and questions about vocabulary or grammatical concepts.
The following sections describe the features and the implementation details of \sysname{}.

\subsection{Interface Design}

\begin{figure*}[t]
    \includegraphics[width=\textwidth]{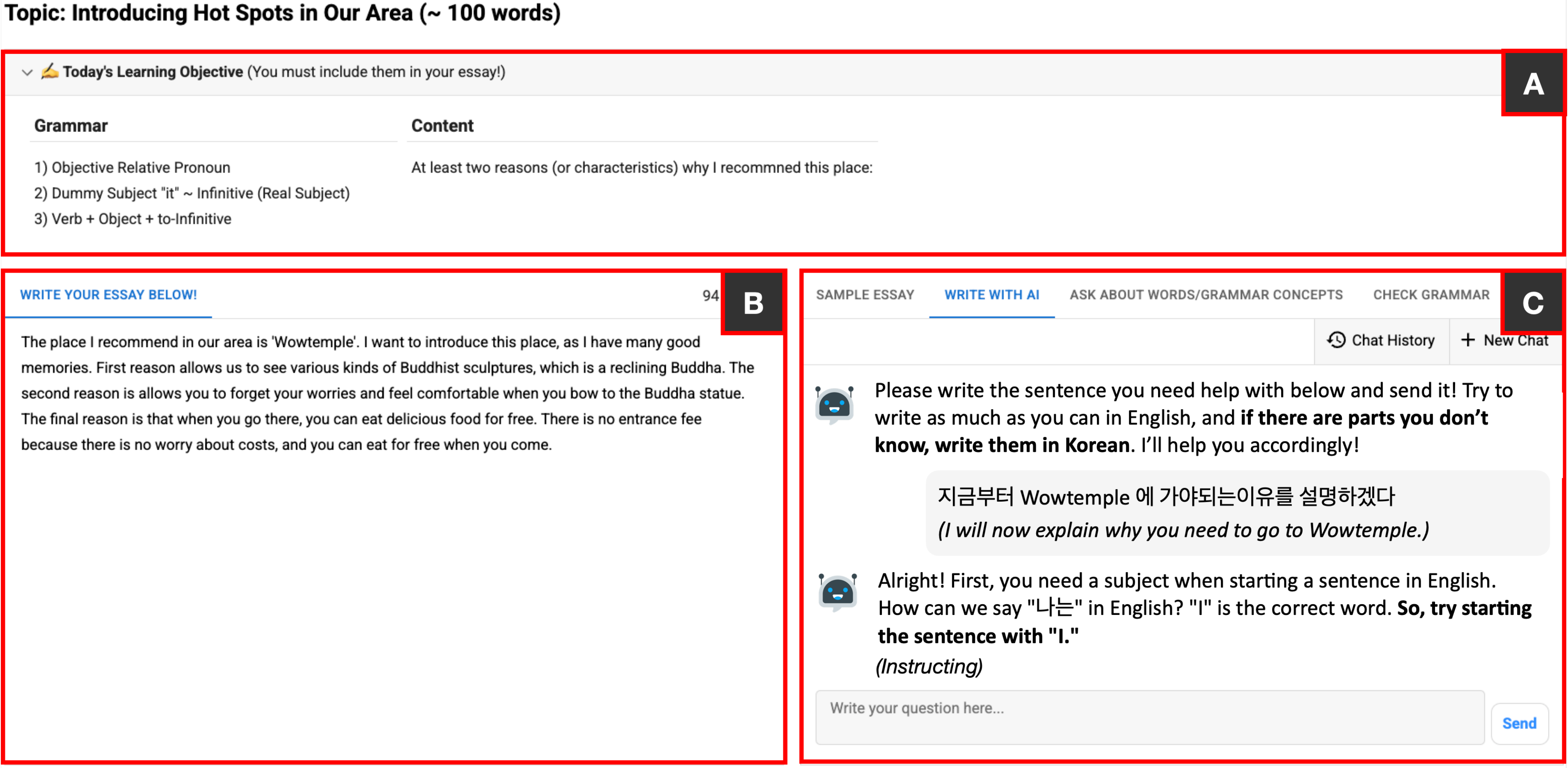}
    \caption{The interface design of \sysname{}. Students can view the lesson objectives in section (A), write their essays in section (B), and interact with the AI chatbot in section (C). From section (C), students can select features they want from the available options in the tab. All instructions are provided in Korean and are translated into English for this figure.}
    \Description{Interface screenshot. The top section (A) shows learning objectives with grammar and content guidelines. The left section (B) is a writing area where students compose their essays. The right section (C) is an AI chatbot window that assists students with English sentence writing.}

    \label{fig:system}
\end{figure*}

We co-designed the system's main features in collaboration with the middle school teacher to align the system with both the students' needs and the class requirements. As most students had little to no prior experience with LLMs, the interface was designed to be simple and approachable. To directly address common difficulties students face in English writing, the system included three core functionalities: (1) collaboratively constructing English sentences from Korean or code-mixed input, (2) providing contextual explanations for new vocabulary and grammar concepts, and (3) guiding students in identifying and revising grammatical errors in individual sentences. These functionalities were drawn from the teacher's prior experience with students' difficulties. 

To reduce the cognitive demands of interacting with an open-ended conversational agent, we organized the system into a set of clearly defined tabs rather than a single unified interface. Our design rationale was informed by the teacher's insight that students at this level were more familiar with task-specific tools such as machine translators and dictionaries, which have clearly defined purposes and interaction patterns. Open-ended conversational interfaces can overwhelm students with low AI literacy, making it difficult to initiate dialogue or formulate appropriate questions \cite{zamfirescu2023why, prather2024widening}. By separating functionalities into clearly labeled tabs with brief guidance prompts, we aimed to lower the barrier to entry and help students understand how to interact with each feature.

The final interface consisted of four tabs, as shown in Figure \ref{fig:system}. The ``Sample Essay'' tab displayed a teacher-curated example essay, a standard component of the existing curriculum, which students could reference during writing. The ``Write with AI'' tab supported collaborative sentence construction, where students could input Korean or code-mixed text and work with the AI to formulate proper English sentences. The ``Ask About Words/Grammar Concepts'' tab enabled students to request contextual explanations for unfamiliar vocabulary or grammatical structures. The ``Check Grammar'' tab guided students in identifying and revising grammatical errors in their sentences. Each tab displayed a brief guidance prompt at the beginning of the session (e.g., ``Please ask me about any words or grammar concepts you find confusing'') to help students understand how to begin their interaction with that particular functionality. During writing activities, students freely navigated between tabs, referring to the guidance prompts and entering their own questions as needed.

In addition to these structural design choices, we also considered how students would express themselves when interacting with the system. We allowed students to enter Korean–English code-mixed input when interacting with the LLM. This design choice is grounded in work showing that code-mixing is a natural and prevalent strategy among beginner-level EFL learners, who frequently draw on their L1 to compensate for vocabulary gaps, clarify meanings, and maintain the flow of communication~\cite{sert2005functions,horasan2014code}. Studies have shown that the strategic use of L1 in the writing process functions as a supportive factor, especially when students face demanding tasks or limited L2 resources \cite{kim2014use, choi2006l1}. Rather than restricting students to English-only input, which would create an artificial interaction environment, we aimed to observe authentic classroom behaviors by allowing students to decide when and to what extent they would code-mix.


\subsection{Prompt Design}
\label{sec:prompt-design}

To align with the teacher's pedagogical requirements, \sysname{} was designed to provide step-by-step guidance rather than giving direct answers. One of the teacher's key requests was to encourage students to actively engage with the writing process by attempting tasks themselves, with the system offering hints and scaffolding incrementally. This design choice reflects prior research showing that simply providing answers---such as in machine translation (MT) tools---without explanations often leads to poor retention and limited learning outcomes \cite{lee2024effects}.

Accordingly, the prompts were designed to elicit the LLM to provide scaffolded support. We provided the LLM with six scaffolding strategies defined by \citet{van2010scaffolding}, along with the current conversation context with the student. The LLM was then instructed to select the most appropriate strategy and generate a step-by-step guide based on the chosen strategy. The details of the scaffolding strategies and the example responses by the LLM are outlined in Table \ref{tab:scaffolding}, and the prompt is shown in Appendix \ref{sec:prompt}.

\begin{table*}[]
\centering
\caption{Scaffolding strategies defined by \citet{van2010scaffolding}. The examples in the table are the translated version of the actual responses provided by the LLM using \sysname{}.}
\Description{Table showing scaffolding strategies used in teaching, with two columns: strategy and description. Strategies include Feeding back, Hints, Instructing, Explaining, Modeling, and Questioning. Each strategy is accompanied by an explanation and an example sentence. For example, `Feeding back' provides information about student performance, while `Modeling' demonstrates how to solve a task by showing examples.}

\label{tab:scaffolding}
\resizebox{\textwidth}{!}{%
\begin{tabular}{l|l}
\hline
\textbf{Scaffolding Strategy} & \textbf{Description} \\ \hline
Feeding back & \begin{tabular}[c]{@{}l@{}}Providing information about the student's performance.\\ e.g.) Great job! Your sentence is perfect. You wrote ``The place where I recommend visiting is the Hoam Art Gallery.'' very well.\end{tabular} \\ \hline
Hints & \begin{tabular}[c]{@{}l@{}}Giving indirect suggestions or clues to help students move forward.\\ e.g.) Can you try using a different word instead of `that'? For example, you could consider expressions like `where' or `in which.'\end{tabular} \\ \hline
Instructing & \begin{tabular}[c]{@{}l@{}}Telling students what to do, explaining how something should be done, and why it is done in that way.\\ e.g.) You need a subject before `would like to.' Let's add `you' and see how the sentence changes.\end{tabular} \\ \hline
Explaining & \begin{tabular}[c]{@{}l@{}}Offering detailed information or clarifications.\\ e.g.) An article is a word that comes before a noun and is mainly used to ...\end{tabular} \\ \hline
Modeling & \begin{tabular}[c]{@{}l@{}}Demonstrating the process of solving a task by showing how to approach similar examples.\\ e.g.) Now, you can also use an infinitive ... Can you try creating one? For example, you can say: ``Feel free to ask any questions.''\end{tabular} \\ \hline
Questioning & \begin{tabular}[c]{@{}l@{}}Asking questions that require students to provide active linguistic and cognitive responses.\\ e.g.) But when the subject is ``you,'' ... it would be "you go," right? Then, what should come after ``When you go there''?\end{tabular} \\ \hline
\end{tabular}%
}
\end{table*}

\section{Method}

We conducted a six-week deployment study at a public middle school in South Korea to examine how K-12 EFL students interact with \sysname{}. 

Our research questions are:
\begin{itemize}
\item \textbf{RQ1:} How do K-12 EFL students with varying proficiency levels engage with \sysname{} during classroom writing activities?
\item \textbf{RQ2:} How do different interaction patterns with \sysname{} influence students' retention and application of grammar and vocabulary knowledge?
\item \textbf{RQ3:} How does integrating LLMs into real-time K-12 EFL writing classrooms affect student engagement, teacher roles, and classroom equity?
\end{itemize}

As a measure of safety, we collaborated with a middle school teacher in Korea, and all the experiments were done under the supervision of the teacher. 
Our study was approved by our institute’s ethics review board prior to deployment.

\subsection{Participants}
We deployed our system to eighth-grade students, who are all between 13 and 14 years old. Since the experiment involved minors, we provided a detailed explanation of our system to their parents and obtained consent only when both the student and their parents agreed to participate. The study was conducted with six classes, each having an average of 26 students. Of the 157 students in total, 133 (56 male, 77 female) consented to participate and share their data for analysis. For students who did not consent, we still provided the same system for optional use, but discarded any data before analysis.

\subsection{Study Setting}
In South Korea, all middle school students are required to learn English as part of the mandatory curriculum. On average, their English proficiency is at the beginner level (A2 in CEFR or below), and most students struggle to express their ideas in complete sentences \cite{lee2024effects}. 

English writing in Korean middle schools is usually conducted as an end-of-unit activity, where students apply newly learned vocabulary and grammar to short compositions. For example, in our experiment, students wrote a 100-word paragraph on the topic ``Introducing Hot Spots in Our Town'' while incorporating specific grammar concepts from the unit, such as the objective relative pronoun. All writing is completed in class under teacher supervision, with guidance provided as needed. Each assignment spans about five class sessions (45 minutes each), taking roughly three weeks to finish.

We deployed \sysname{} during two writing exercises, one in spring 2024 and another in fall 2024 (see Appendix \ref{sec:study_details} for details on topics and objectives). The Korean academic year runs from spring to fall. For these deployments, we used gpt-4-turbo-2024-04-09 in the first semester and gpt-4o-mini-2024-07-18 in the second, as these were the most up-to-date and cost-effective models available at the time.

\subsection{Observation Study}
To complement the system logs and written outputs, we conducted a naturalistic observation study \cite{angrosino2016naturalistic} to better understand how students engaged with \sysname{} during in-class writing activities. Observations were carried out across three class sessions (45 minutes each). Due to the privacy regulations and ethical considerations, video or audio recordings were not permitted. Instead, the authors took detailed field notes throughout the sessions, focusing on students' collaborative behaviors (e.g., discussing outputs with peers), and moments when they asked for clarification or assistance from the teacher. The notes were later analyzed for recurring themes across sessions.

\subsection{Written Assessment}
To measure learning outcomes, a written assessment on the same topic was conducted 2–3 weeks after the writing session as part of the regular class. In this assessment, students wrote the essay with the same topic independently, without any support from the teacher or AI tools. We collected the final essays and the corresponding scores of students who consented to share their data to evaluate the effectiveness of AI support.

Based on the students' assessment scores, participants were categorized into three performance levels: high-, middle-, and low-performing groups. The essays were graded using an analytic rubric adapted from \citet{jacobs1981testing}. The rubric covered four domains: content, organization, vocabulary, and language use. In this rubric, the language use domain captured grammatical accuracy (e.g., tense, subject-verb agreement, and basic mechanics) as well as the appropriateness of sentence structures. Each domain was scored on a 5-point scale, and the grading was conducted by a middle school English teacher with an MA in English education. Students scoring 19–20 were labeled high-performing, 13–18 middle-performing, and 12 or below low-performing. Detailed rubric is outlined in Appendix \ref{sec:rubric}. Also, the distribution of the final scores of the students is shown in Appendix \ref{sec:score}

\subsection{Analysis}
We conducted both quantitative and qualitative analyses to examine how students used \sysname{} in their classroom writing activities and how it influenced their writing performance. For quantitative analysis, we conducted a descriptive statistical analysis to understand the usage patterns of students and their overall participation. 

To investigate the interaction patterns that emerge when K-12 students engage with an LLM, we conducted a qualitative coding of student queries to categorize the types of questions students asked. We first randomly selected five students from each performance group (high-, middle-, and low-performing) and sampled their queries. This resulted in a dataset of 1,386 queries, representing 9.3\% of the total. Then, two of the authors familiarized themselves with the sampled queries, and independently open-coded the data to identify recurring patterns. Through discussion, the authors generated a preliminary codebook, which was further validated against the remaining data. The final version of the codebook is presented in Table~\ref{tab:codebook}. Using this codebook, two authors independently coded the remaining queries. Inter-rater reliability was assessed using Cohen's Kappa, resulting in a value of 0.77, indicating substantial agreement. Any disagreements were resolved through discussion between the two authors until consensus was reached.

\begin{table*}[]
\centering
\caption{The final codebook used for qualitative coding, along with definitions and examples drawn from the analysis results.}

\resizebox{0.85\textwidth}{!}{%
\begin{tabular}{l|l}
\hline
\multicolumn{1}{c|}{\textbf{Category}} & \multicolumn{1}{c}{\textbf{Description}} \\ \hline
\textbf{Grammar Concept} & \begin{tabular}[c]{@{}l@{}}Questions about grammar rules and differences.\\ Example: ``Should I use `Korea have' or `Korea has'?''\end{tabular} \\ \hline
\textbf{Vocabulary Usage} & \begin{tabular}[c]{@{}l@{}}Inquiries about word meanings and usage.\\ Example: ``What's the difference between `scale' and `size'?''\end{tabular} \\ \hline
\textbf{Vocabulary Translation} & \begin{tabular}[c]{@{}l@{}}Requests to translate specific words.\\ Example: ``What is `흔히' in English?''\end{tabular} \\ \hline
\textbf{Improvement Seeking} & \begin{tabular}[c]{@{}l@{}}Seeking better expressions beyond correctness.\\ Example: ``How can I make this sentence longer?''\end{tabular} \\ \hline
\textbf{Validation Seeking} & \begin{tabular}[c]{@{}l@{}}Checking the correctness of an answer.\\ Example: ``So I shouldn't use `the' here?''\end{tabular} \\ \hline
\textbf{Sentence Revision} & \begin{tabular}[c]{@{}l@{}}Providing an English sentence for correction.\\ Example: ``It is the reason for me to introduce this place...''\end{tabular} \\ \hline
\textbf{Paragraph Revision} & \begin{tabular}[c]{@{}l@{}}Providing a paragraph for correction.\\ Example: ``First, the place I recommend is Wow Temple... Second, ...''\end{tabular} \\ \hline
\textbf{Partial Sentence Generation} & \begin{tabular}[c]{@{}l@{}}Generating missing parts of a sentence.\\ Example: ``When you visit Wow Temple, you can see 불상의 다양한 조각상.''\end{tabular} \\ \hline
\textbf{Full Sentence Generation} & \begin{tabular}[c]{@{}l@{}}Requesting to generate the whole sentence in English.\\ Example: ``Translate: 이 장소를 추천하는 이유는...''\end{tabular} \\ \hline
\textbf{Interactive Response} & \begin{tabular}[c]{@{}l@{}}Non-question responses reacting to LLM's feedback\\ Example: ``Oh, that makes sense!'' / ``I don't get it.''\end{tabular} \\ \hline
\end{tabular}
}
\Description{Table showing the final codebook for qualitative coding, with two columns: category and description. Categories include Grammar Concept, Vocabulary Usage, Vocabulary Translation, Improvement Seeking, Validation Seeking, Sentence Revision, Paragraph Revision, Partial Sentence Generation, Full Sentence Generation, and Interactive Response. Each category is defined with a short description and illustrated with an example sentence.}
\label{tab:codebook}
\end{table*}

Furthermore, to assess the efficacy of LLM-based scaffolding, we used GPT-4o, the best-performing model at the time of analysis, to extract the final English sentence written by students from each conversation thread. The analysis focuses on student interaction data from the fall semester, and on the use of the \textit{sentence construction} feature. We then assessed the grammatical correctness of these sentences using GPT-4o, as previous studies demonstrated high accuracy of LLMs in detecting grammar errors \cite{coyne2023analyzing}. Finally, we compared the extracted sentences with students’ final essays from the written assessment to examine whether the sentences were correctly recalled. The prompts used for this analysis are provided in Appendix \ref{sec:prompt}.

\section{Result}

In this section, we first report how students with different proficiency levels engaged with \sysname{} during in-class writing sessions, drawing on both descriptive statistics and thematic analysis of the system log data (RQ1). We then report written assessment outcomes along with statistical analysis to explore how interaction patterns affect students' retention and application of grammar and vocabulary (RQ2). Finally, we present insights from classroom observations to report the effects of integrating LLMs into real-time K-12 EFL writing classrooms, including student engagement, teacher roles, and classroom equity (RQ3).

\subsection{RQ1: K-12 EFL Students' Engagement with \sysname{}}

\subsubsection{Descriptive Statistics}
\begin{figure*}[t]
    \includegraphics[width=0.7\textwidth]{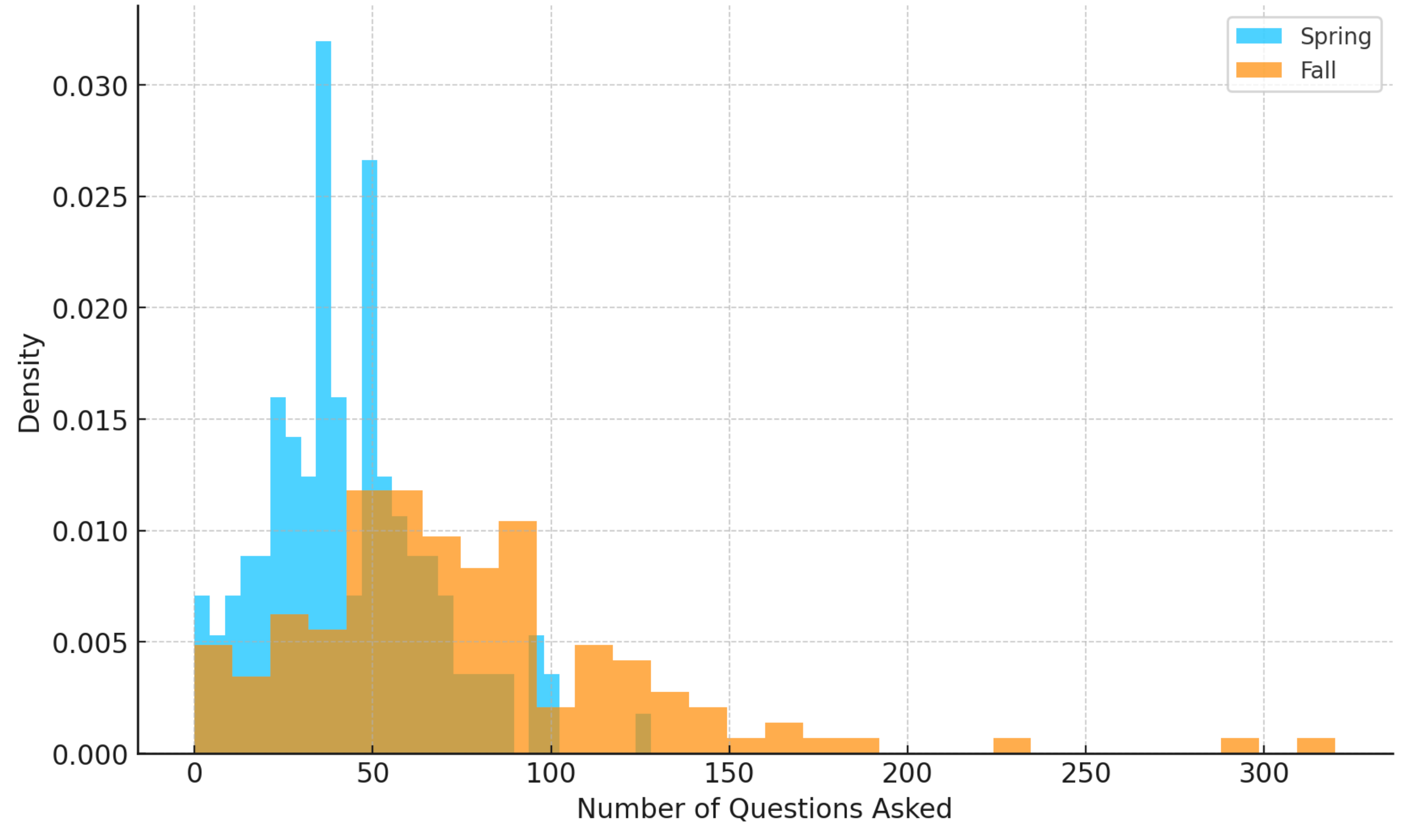}
    \caption{Distribution of the number of query-response pairs between students and the LLM during each semester.}
    \Description{Histogram comparing the distribution of the number of questions asked by students to the LLM in two semesters. The x-axis shows the number of questions, and the y-axis shows density. The blue bars represent the spring semester, and the orange bars represent the fall semester. Spring shows a higher concentration of students asking fewer than 60 questions, while fall has a wider spread, with some students asking more than 150 questions.}

    \label{fig:frequency}
\end{figure*}

During the six-week deployment study, students sent a total of 14,863 messages to the LLM. On average, each writing session included 59.5 messages sent per student ($SD=42.1$, $min=0$, $max=320$). Figure \ref{fig:frequency} illustrates the distribution of the number of messages sent per student. Notably, students asked significantly more questions in the fall semester, suggesting that as they became more familiar with the system, their engagement increased.

A Kruskal–Wallis test was conducted to compare user activity across the high-, middle-, and low-performing groups. The analysis revealed a significant difference in the number of messages per conversation thread ($H(2)=55.39$, $p<.001$). Post-hoc Dunn’s tests with Bonferroni correction indicated that conversation threads in the high-performing group were significantly shorter than those in both the middle-performing group ($p_{\text{adj}}<.001$) and the low-performing group ($p_{\text{adj}}<.001$). This result suggests that high-performing students resolved their queries with fewer exchanges, reflecting more efficient interactions within each thread.

In terms of feature usage, students most frequently used the \textit{sentence construction} feature (11,476 queries, 72.2\%), followed by \textit{asking vocabulary and grammar concepts} (3,075 queries, 19.3\%) and \textit{grammar revision} (1,351 queries, 8.5\%). A Chi-square test was performed to examine the relationship between student groups and their use of different task types. The analysis revealed a statistically significant association $\chi^2(4, N=14,863)=211.10$, $p<.001$.
Although \textit{sentence construction} remained the dominant task type in all groups, post-hoc Z-tests with Bonferroni correction revealed that the high-performing group made significantly greater use of the \textit{asking vocabulary and grammar concepts} and \textit{grammar revision} features compared to the middle- and low-performing groups ($p<.001$).

Furthermore, an analysis of the interaction timeline revealed that students, regardless of their group, predominantly used the grammar revision feature during the latter half of their conversations (77.4\% of its use). Given that high-performing students used this feature significantly more overall, this suggests they were able to advance to the revision stages more quickly and had greater opportunities to engage with the LLM to learn the grammatical concepts they had initially struggled with. These findings highlight that time constraints in real-time classroom settings may influence students' engagement and patterns of feature use.

Students asked queries in English (6,200, 42.08\%), Korean (3,843, 26.08\%), and in a code-mixed format (4,693, 31.84\%). 
Among the code-mixed queries, 4,153 involved intra-sentential code-mixing, typically occurring when students asked about the specific meaning of an English word in Korean, or wrote unknown parts in Korean and requested the model to generate a complete English sentence. Inter-sentential code-mixing occurred when students had a specific request to the model regarding the full English sentence they wrote, such as ``The place taught people to appreciate valuable history and gave a strong appreciation for our cultural heritage. 이렇게 바꾸면 자연스러워?'' (Is this natural now?)

A Chi-square test revealed statistically significant differences in language usage patterns among high-, middle-, and low-performing groups $\chi^2(4, N=14,736)=53.82$, $p<.001$. To pinpoint the source of these variations, post-hoc Z-tests with Bonferroni correction were performed. Contrary to expectations, the middle-performance group used English most frequently (47.1\%), significantly more than both the high- (40.0\%) and low-performance (42.0\%) groups ($p<.001$). This group also demonstrated a significantly lower use of code-mixing (28.1\%) compared to the high- (32.6\%) and low-performance (33.3\%) students ($p<.001$). Regarding Korean usage, while the overall rates were similar, a statistically significant difference was found between the high- (27.4\%) and low-performance (24.7\%) groups ($p<.05$).

\subsubsection{Qualitative Analysis}

\paragraph{Question Types}
\label{sec:question_type}

\begin{figure*}[t]
    \includegraphics[width=0.8\textwidth]{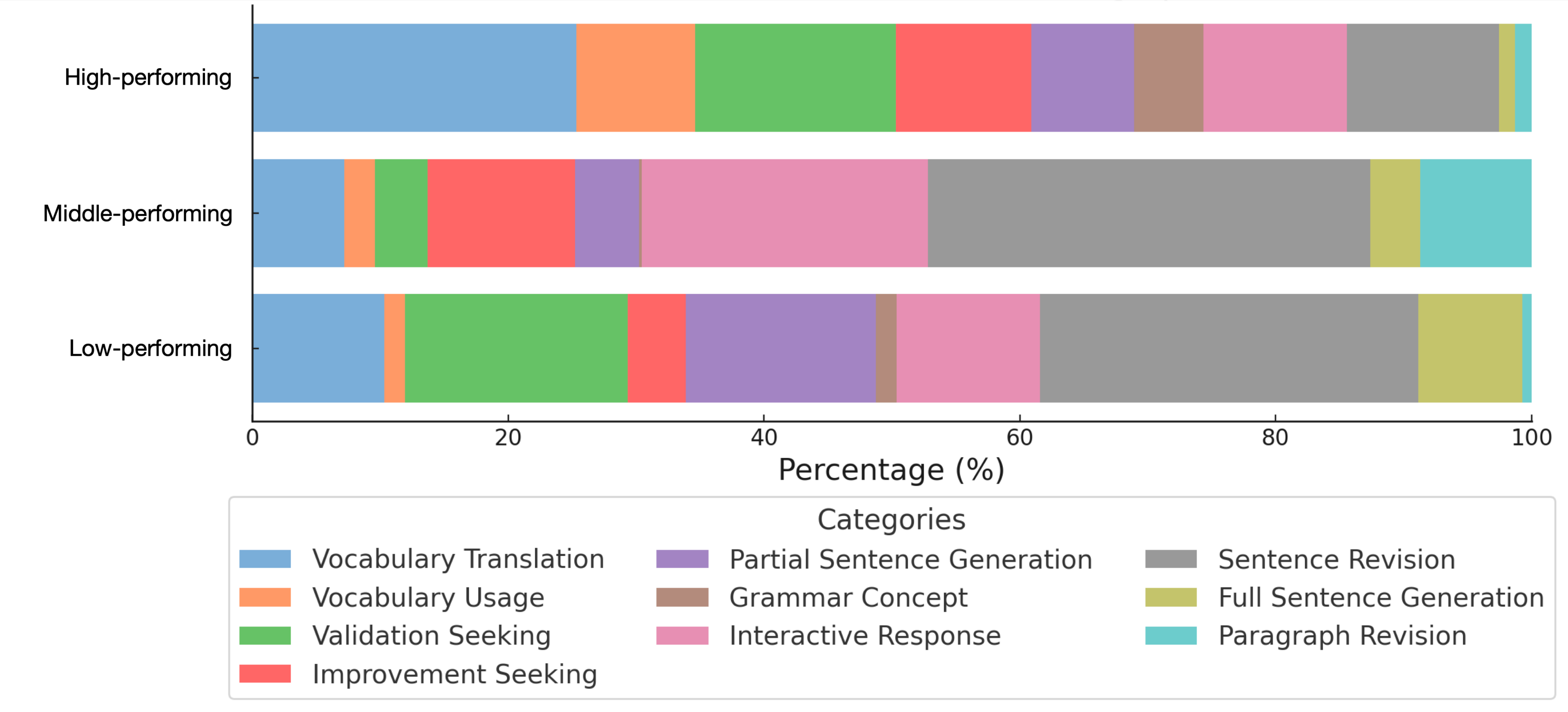}
    \caption{Distribution of the question types asked across different student performance levels.}
    \Description{Stacked bar chart showing the distribution of question types asked by students at different performance levels: high-, middle-, and low-performing. The x-axis represents percentage, and each bar is divided into categories such as Vocabulary Translation, Vocabulary Usage, Validation Seeking, Improvement Seeking, Partial Sentence Generation, Grammar Concept, Interactive Response, Sentence Revision, Full Sentence Generation, and Paragraph Revision. High-performing students show a more balanced mix, middle- and low-performing students ask many sentence revision questions.}
    \label{fig:question_type}
\end{figure*}

Figure \ref{fig:question_type} presents the frequency distribution of question types across different student proficiency levels. A notable distinction is that students in the middle- and low-performing groups exhibit a significantly higher frequency of \textit{sentence revision} requests compared to those in the high-performing group. Specifically, sentence revision accounts for 34.6\% of all questions in the middle-performing group and 29.5\% in the low-performing group, whereas it only constitutes 11.9\% in the high-performing group. Similarly, paragraph revision is more common among middle-performing students (8.7\%) when compared to high-performing (1.3\%) students. 

Sentence and paragraph revision requests from lower-proficiency students originate from the process of incorporating feedback from the LLM while constructing sentences. An overwhelming 327 out of 366 revision requests were given during the use of \textit{sentence construction} feature. These findings suggest that lower-proficiency students engage more in iterative revision processes rather than generating fully-formed sentences independently. This interpretation is further supported by the higher average number of queries per thread in sentence construction tasks, with high-performing students averaging 4.56 queries per thread, middle-performing students at 5.73, and low-performing students at 6.02.

Additionally, students in the high-performing group display a stronger tendency toward \textit{vocabulary translation} (25.3\%) and \textit{validation seeking} (15.7\%), indicating a focus on precise word choice and confirming correctness. In contrast, the low-performing group shows a higher frequency of \textit{full sentence generation} (8.2\%) and \textit{partial sentence generation} (14.9\%), which implies their struggles with independently constructing complete sentences. These differences can be interpreted through Bloom's Taxonomy \cite{krathwohl2002revision}. High-performing students delegate lower-order tasks like vocabulary translation (Remembering) and validation seeking (Understanding) to the LLM. This frees them to concentrate on higher-order thinking, such as `Applying' the learned grammatical concepts and `Creating' their own work. Conversely, low-performing students use the LLM for sentence generation, outsourcing the core skills of `Applying' rules and `Creating' text. 


For students' \textit{interactive responses}, we conducted a further qualitative analysis focusing on the negative affective behaviors they exhibited. These responses were categorized into three affective expressions: gratitude (e.g., ``Thanks!''), annoyance (e.g., ``Give me the right answer!!''), confusion (e.g., ``I don’t get it. You do it.''), and indifference (e.g., ``Whatever.'').

The results show a notable difference in the frequency of queries expressing confusion or annoyance. On average, the high-performing group exhibited these affective behaviors only 3.8 times throughout each writing session, while the middle- and low-performing groups showed 12.2 and 11.4 times, respectively. Additionally, the contexts in which negative affective behaviors occurred showed meaningful differences. High-proficiency students expressed annoyance or confusion primarily when the LLM failed to provide responses aligned with their precise intents. For instance, a student displayed confusion when the LLM provided an indirect hint rather than a direct answer to the query, ``What is a peak in English?'' In contrast, low-proficiency students exhibited annoyance or confusion when they became overwhelmed by the LLM’s step-by-step feedback. The most frequent case involved students repeatedly responding with ``I don't know'' to the LLM's comments, even when the model was actively guiding them toward a solution. This result highlights the importance of adopting different interaction strategies with LLMs across proficiency levels to promote students' satisfaction and positive engagement. 

Furthermore, among the middle- and low-performance groups, the majority of negative affective behaviors (69.2\%) emerged in the latter half of the interaction. In contrast, the high-performance group displayed a more balanced pattern, with affective responses distributed evenly across the session (52.5\% in the latter half). One possible explanation is that as unanswered questions accumulated, some students may have begun to feel discouraged, which could have reduced their motivation and fostered reliance on the system rather than actively working through the hints. Another possibility is that time pressure toward the end of the session led to frustration, and the absence of direct answers in such moments may have further contributed to the rise in negative affective behaviors.

\begin{figure*}[t]
    \includegraphics[width=\textwidth]{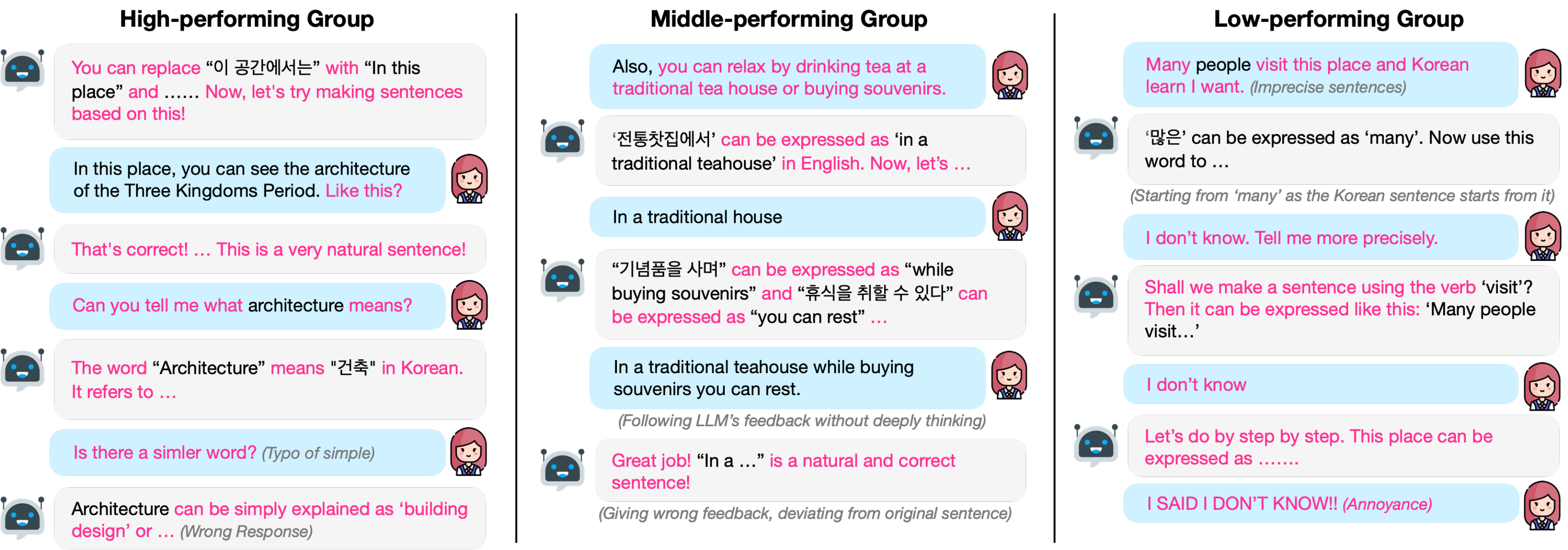}
    \caption{Example interaction patterns that emerged from the use of \sysname{}. The interactions are drawn from actual data of students from high-, middle-, and low-performance groups. Text in pink was originally written in Korean and translated into English for reader's clarity.}
    \Description{Figure showing example interaction patterns between students and the LLM across three performance groups: high-, middle-, and low-performing. The left column (high-performing) shows constructive exchanges, including word meaning explanations, natural sentence feedback, and simple clarifications. The middle column (middle-performing) includes partial corrections, natural sentence confirmations, and occasional deviations from the original sentence. The right column (low-performing) shows struggles with vague expressions, repeated uncertainty, and frustration responses.}
    \label{fig:interaction}
\end{figure*}

\subsection{RQ2: \sysname{}'s Effects on Grammar and Vocabulary Learning}
\label{sec:rq2}



During the fall semester, students constructed a total of 632 sentences with the assistance of an LLM. Of these, 309 sentences (48.9\% of the total) were incorporated verbatim into the students' final essays, with 82.3\% of them being grammatically correct. This trend was also observed in the low-performing group, where 86.2\% of the sentences constructed with the LLM were grammatically accurate. However, while students in the high-performing group correctly incorporated 65.8\% of these sentences into their final written assessments, students in the middle- and low-performing groups only did so 37.8\% and 14.0\% of the time, respectively. These findings suggest that, regardless of proficiency level, the use of an LLM can serve as an effective tool for improving students’ grammatical accuracy during in-class activities. However, the degree to which students retain and appropriately transfer these sentences into their final written work appears to differ across performance groups.

Errors from LLMs mostly originated from students' code-switched sentences containing imprecise elements, such as omitted subjects or predicates. These issues largely originated from the grammatical differences between Korean and English. For example, while subject omission is common in Korean \cite{park2013statistical}, it is not acceptable in English. When students asked the LLM to correct sentences written in Korean syntax without accounting for these differences, the LLM sometimes misinterpreted the intended meaning, producing sentences that were grammatically incorrect or semantically altered.

Another source of error arose from differences in word order between Korean and English. When these structural differences were not considered, the LLM sometimes provided scaffolding that resulted in awkward constructions. For example, a student wrote in Korean, ``Many people visit this place and Korean I want,'' which is grammatically correct. Instead of guiding the student on how to restructure the sentence in English properly, the LLM focused on translating individual elements, beginning with how to write ``Many'' in English. which could lead the student to produce sentences with incorrect or unnatural word order. This approach by the LLM led the student to produce sentences with incorrect or unnatural word order.

Finally, errors also occurred frequently when students attempted to incorporate in-class requirements into their sentences. For instance, the use of an object relative pronoun was a mandatory requirement assigned by the teacher, and students often asked the LLM to include it even in sentences where it did not naturally fit. In such cases, the LLM generated sentences with unnatural structures instead of indicating that the request was inappropriate. An example is the sentence ``I want to introduce Poeun art hole,'' which the LLM corrected as ``I want to introduce a place that is Poeun art hole'' after the request from the student.



\subsection{RQ3: Impact of Real-Time LLM Integration on Student Engagement, Teacher Roles, and Classroom Equity}

This section investigates the integration of LLMs into real-time K–12 EFL writing classrooms, focusing on how they reshape classroom dynamics. Drawing on classroom observations and a follow-up teacher interview, we analyze both the benefits and challenges of LLM use in supporting real-time learning.

\subsubsection{Roles of LLMs, Peers, and Teachers in Real-Time EFL Classrooms}

Our classroom observations revealed clear advantages in terms of scalability. Students often asked relatively simple questions, such as how to spell specific words in English or when and how to use basic grammatical concepts (e.g., articles). In the past, the teacher would have to answer these one by one and demonstrate how to apply the answers in context. With the introduction of the \sysname{}, however, the teacher could redirect students by saying, ``Try asking the LLM like this,'' which enabled faster responses to repetitive, low-level questions. This finding was confirmed with the follow-up interview, where the teacher noted ``I had more time to walk around the classroom and check students’ progress, since I no longer needed to respond to every simple question.''

Despite these benefits, there was a key limitation: the roles of the teacher and the LLM were not clearly defined. We observed that students would often ask questions to the teacher first, even with simple questions that the LLM could have easily handled. This pattern led the teacher to constantly step in and redirect them by saying, ``Have you tried asking the model first?''  This highlights the importance of clearly defining the roles of the teacher and the LLM in a real-time class, and guiding students on how to use both effectively and complementarily.

Another important observation was the emergence of new forms of peer interaction. During classroom observations, students occasionally discussed about the responses generated by the LLM. They compared outputs, commented on which answers seemed more useful, and sometimes worked together to interpret how to apply them in their own work. They also shared their excitement in conversing with the LLM, for instance, by saying ``It gave me a compliment!'' These episodes show that integrating LLM in real-time classroom can introduce a new dimension of social interaction in EFL classrooms.

However, the overall peer interaction was not particularly active. Many of the questions that students previously directed toward more proficient classmates were now asked to the LLM instead. While this change increased the accessibility to question-asking, it also reduced opportunities for collaborative learning. Peer learning is important in EFL education; students with lower proficiency can learn by articulating what they find difficult, while students with higher proficiency can build confidence and deepen their understanding of grammar by helping their peers \cite{won2017effect}. Thus, the integration of the LLM needs to be carefully balanced so that it supports efficiency without undermining the valuable role of peer learning.

\subsubsection{Influence of LLM on Classroom Equity}

The introduction of \sysname{} had mixed effects on classroom equity. On the one hand, it expanded access to learning tasks. Students with lower proficiency, who might previously have struggled to complete assignments on their own, were now able to finish their work with the support of the AI. In the follow-up interview, the teacher mentioned ``Previously, the gap between the top and bottom students was very wide, but with the help of the LLM, many students managed to complete their tasks, so more students ended up at an intermediate level.'' This suggests that the introduction of LLM in a real-time classroom can help reducing gaps in task completion and give students a greater sense of accomplishment.

On the other hand, the AI also introduced new challenges for the equitable use of teacher resources. In every observed class, only four to five students out of 26 dominated the teacher's attention by asking frequent questions. While these few active students continued to seek clarification from the teacher despite the presence of the LLM, more introverted students asked even fewer questions than before, instead relying on the LLM as a more approachable alternative. The teacher interview confirmed this trend, by saying ``students who asked few questions before have been asking even fewer.'' This dynamic highlights the need to consider how LLMs influence classroom participation and teacher-student interactions.

Also, the introduction of \sysname{} made it more challenging for the instructor to identify common student difficulties, as questions were now split between the teacher and the AI assistant. A common in-class practice for the teacher was to observe recurring student questions, and when a particular concept proved challenging for multiple students, the teacher paused the writing activity to provide a brief lesson to the entire class. With fewer students asking questions to the teacher, it became more difficult to identify common errors and deliver timely lessons on those issues.

\section{Discussion}

In this study, we investigated how K-12 EFL students interact with LLMs in a real-time classroom writing context. To this end, we designed and deployed \sysname{} as a technology probe to collect authentic AI-mediated writing experiences and envision future design considerations. In this section, based on the findings of the research probe, we discuss the design implications for using LLM as a more effective and equitable tool in real-time educational settings.

\subsection{Learner-Centered Design: From Task Completion to Long-Term Learning}

\subsubsection{Provide Syntax-Aware, Cross-Lingual Scaffolding}
Our study revealed that students' and LLMs' errors often originate from L1 interference, such as applying Korean grammatical rules (e.g., subject omission) to English sentences. The LLM, failing to understand this underlying cause, often provided corrections that were either unhelpful or generated new, unnatural sentences.
In contrast, EFL teachers, fluent in both languages, strategically leverage students’ L1 in the classroom. By understanding common errors that arise from syntactic differences between Korean and English, teachers can more accurately identify and address misunderstandings while reducing student anxiety \cite{wolfersberger2003l1, uzawa1996second}.

Prior learning analytics and automated writing evaluation systems for writing have largely treated learner input as monolingual and focused on surface error types or holistic text quality \cite{wang2013exploring,liaqat2018towards}, making it difficult to capture cross-lingual interference patterns.
Future systems should be designed with an awareness of common cross-linguistic challenges, incorporating the tacit knowledge of EFL educators on how to address them. Rather than simply translating, LLMs should recognize common learner errors rooted in L1 syntax and guide students in restructuring their ideas according to the target language.

\subsubsection{Implement Dynamic, Proficiency-Aware Scaffolding}
Our findings from the deployment study challenge the prevailing assumptions in existing research on LLMs in educational settings that focus on the non-disclosure of answers as a pedagogical strategy \cite{kazemitabaar2024codeaid, lee2023learning, heickal2024generating}. We reveal that this approach may not be appropriate for students with lower levels of proficiency and low motivation to learn, as withholding answers can be counterproductive, potentially leading to frustration, disengagement, and repeated queries. This issue becomes more pronounced in real-time classroom settings, where time constraints significantly impact student-LLM interactions. As the non-disclosure of answers forces students to follow a longer, more complex path, it may hinder their learning when they cannot engage with the given feedback at their own pace due to time limitations. Therefore, LLM-based scaffolding approaches should consider not only students' proficiency levels, but also multifaceted aspects such as their internal motivation, ability, and time required to process feedback, and time constraints given in real-time classrooms.

\subsection{Classroom-Centered Design: Supporting Teachers and Fostering Equity}

\subsubsection{Clearly Define and Communicate the Roles of the LLM and the Teacher}
The ambiguity of LLM's role in a real-time classroom led to inefficiencies, with students asking the teacher simple questions that LLM could handle, and the teacher constantly redirecting them. To better leverage the limited teacher resources, both the LLM's role and the teacher's role must be explicitly defined. The LLM should be positioned as the first responder for lower-order queries (e.g., spelling, vocabulary, basic grammar), freeing up the teacher to focus on higher-order concerns, such as content development, critical thinking, and personalized instruction. 
This approach aligns with prior work on classroom orchestration tools and intelligent tutoring systems, which demonstrate that offloading routine instructional tasks to AI systems allows teachers to dedicate their time to more complex pedagogical activities \cite{holstein2017intelligent, van2019orchestration}.
Additionally, the LLM should be self-aware of its role, responding to questions within its scope and redirecting students to the teacher for more complex, higher-order issues.

\subsubsection{Enhance Teacher Awareness}
While the LLM provided scalable support, it also made it more difficult for the teacher to identify the common struggles of the students, a task previously accomplished by listening to recurring questions. To counteract this, future systems must include a real-time teacher-facing dashboard. 
Research on the use of teacher dashboards for summarizing students’ interactions with LLMs exists, but it has largely focused on asynchronous settings \cite{kim2024llm}. A real-time dashboard should provide analytics on common student errors, learning progress, and engagement levels, enabling teachers to develop an accurate understanding of their students’ needs, rather than being influenced by a small number of extroverted students who ask questions more frequently. Such a tool would not only empower teachers to make data-informed pedagogical decisions in real time but also support the design of equitable learning experiences, ensuring that all students’ challenges and achievements are visible and addressed.

\subsubsection{Mitigate Inequity and Promote Collaboration}
Existing literature has highlighted the supportive role LLMs play for shy or introverted students, offering them a space to engage without the anxiety of speaking up in class \cite{park2023thinking, warman2023empowering}. However, this dynamic does not translate seamlessly into real-time classroom environments. Our observations reveal that while active students continued to seek clarification from the teacher, the presence of the LLM led to a decrease in questions from other students, who found it easier to approach the LLM without anxiety. As a result, more proactive students received increased attention from the instructor, while less engaged students became more reliant on the LLM, leaving them more vulnerable to potential misinformation. Thus, a critical design guideline is to design a system that alerts the instructor with students who are becoming overly reliant on the AI, enabling timely and equitable human intervention.

Furthermore, we found that the tool's efficiency came at the cost of valuable peer-learning opportunities, as it redirected questions that students might have previously posed to their classmates to the AI.  
However, we observed new opportunities for novel forms of collaboration, as students occasionally compared the LLM's outputs and discussed its suggestions. Future systems should consider how LLMs can facilitate collaborative engagement among students, encouraging them to discuss and critique the AI's responses with peers. Such interactions not only support social learning but may also promote more critical and reflective Human-AI engagement, helping students evaluate the tool's suggestions rather than simply accepting them.

\section{Limitation}

This study presents several limitations and opportunities for future research. The study was conducted with eighth-grade students at a single public middle school in South Korea. Therefore, the findings may not be generalizable to students of different age groups (e.g., elementary or high school), in different educational settings (e.g., private schools, non-Asian contexts), or from different cultural backgrounds. In particular, the interaction patterns observed, which were influenced by the linguistic characteristics of Korean and the local curriculum, may differ in different settings. 

Furthermore, we identified meaningful patterns in students' LLM interactions. Due to the extensive size of the collected interaction logs (14,863 utterances), a complete manual review was impractical. Instead, we analyzed a randomly sampled subset comprising 9.3\% of the total data to identify overarching trends. While this approach enabled us to identify overarching trends, it suggests that our findings may not capture the full spectrum of interaction patterns.

Lastly, the findings of this study primarily rely on the teacher's perspectives and the authors' observations. Due to the nature of the classroom deployment, incorporating detailed surveys or interviews---such as those assessing students' satisfaction with each instance of LLM-generated feedback---was not feasible, as such interventions could have disrupted the learning process ~\cite{han2023recipe}. While student satisfaction is not a direct measure of learning outcomes ~\cite{wang2025productive}, fostering greater student agency in their interactions with LLMs and their overall approach to learning appears to be a crucial factor. Future research should explore methods to better integrate student perspectives without compromising the natural flow of classroom activities.

\section{Conclusion}

This paper examined the integration of LLM-based writing support in real-time K–12 EFL classrooms. While LLMs improved grammatical accuracy and offered scalable assistance, their step-by-step scaffolding sometimes hindered lower-proficiency students under time constraints, contributing to demotivation and over-reliance. These students also tended to outsource higher-order cognitive tasks, whereas higher-performing students delegated mostly lower-level tasks, such as vocabulary queries. Classroom observations further revealed social costs. Introverted students asked fewer questions to the teacher, peer learning interactions declined, and struggling students’ difficulties were masked by AI-polished outputs. Taken together, these findings suggest that LLM support can unintentionally widen proficiency gaps and reshape both individual learning and classroom dynamics. Future implementations should focus on promoting equitable participation, preserving teacher awareness, and preserving the social interactions essential for effective learning.


\begin{acks}
This work was supported by Institute of Information \& communications Technology Planning \& Evaluation (IITP) grant funded by the Korea government(MSIT) (No. RS-2024-00443251, Accurate and Safe Multimodal, Multilingual Personalized AI Tutors).
This work was supported by the National Research Foundation of Korea (NRF) grant funded by the Korea government (MSIT) (No.RS-2024-00406715).
This work was supported by Elice.
\end{acks}

\bibliographystyle{ACM-Reference-Format}
\bibliography{references}

\clearpage
\appendix
\section{Prompts}
\label{sec:prompt}

\subsection{Response Generation Prompt}
As discussed in Section \ref{sec:prompt}, we use the CoT prompting method to prompt the LLM to choose the most appropriate scaffolding technique and then generate the scaffolded response to the student. Below is the exact prompt we use in \sysname{}. Note that for the three features, we use a slightly different prompt by changing their role description.

\begin{framed}
\noindent
Here are six scaffolding techniques commonly used by instructors:
\\
\\
1. Feeding back: Providing information regarding the student's performance to the student him/herself. Feedback should not involve direct answers to the task, but should be done with giving explanations and providing similar examples.\\
2. Hints: Provision of clues or suggestions by the teacher to help the student go forward. Teachers must give indirect hints only. More direct hint should only be provided when students say that they don't know the answer.\\
3. Instructing: Teacher telling the students what to do or explanation of how something must be done and why.\\
4. Explaining: Provision of more detailed information or clarification by the teacher. Explanations should involve similar examples (but not the answer of the task).\\
5. Modeling: The process of offering behavior for imitation. This is often done by showing the process of task solving with similar example. Solution to the task should not be given to the students in the process.\\
6. Questioning: Asking students questions that require an active linguistic and cognitive answer.\\
\\
You are an English teacher, teaching an EFL student with a basic level of English. Your task is to guide the student in Korean, to write the correct sentence in English. You should not provide the answer at once, and help the student step-by-step using the scaffolding techniques above. If a student continuously struggle in answering, provide the answer with explanations. If the student successfully wrote the sentence, then praise the student for their job and finish the session by saying ``물어보고 싶은게 있으시면 언제든지 다시 물어봐주세요!''
\\
\\
Please look at the chat record and determine the most suitable scaffolding method. Using that scaffolding technique, return the next message of the teacher in Korean.
\\
Chat record: {Chat Record}
\end{framed}

\subsection{Analysis Prompt}
In this section, we provide the prompt used for analyzing the responses of \sysname{}. 

\begin{framed}
\noindent
Extract the final sentence written by the student from a conversation between a student and a GPT tutor. If the student’s final response is not a complete sentence in English, return an empty string. Additionally, check whether the sentence contains any grammatical errors. If it is grammatically correct, set correctness to true; otherwise, set it to false. 
\\\\
Conversation: {Conversation Thread}
\end{framed}

\section{Study Details}
\label{sec:study_details}
Table \ref{tab:study-details} shows the topics and lesson objectives used for the deployment study done at 2024 spring and fall.

\begin{table*}[]
\centering
\caption{Topic and the lesson objectives used for the deployment study.}
\Description{Table listing topics and lesson objectives for the deployment study. For Spring 2024, the topic is `The best eco-friendly invention,' with objectives including describing two or more characteristics, using subject relative pronouns, and passive voice. For Fall 2024, the topic is `Introducing hot spots in our area,' with objectives including giving two or more reasons for recommendation, using objective relative pronouns, anticipatory it + to-infinitive, and verb + object + to-infinitive structures.}
\label{tab:study-details}
\begin{tabular}{|l|l|l|}
\hline
 & Topic & Lesson Objective \\ \hline
2024 Spring & The best eco-friendly invention & \begin{tabular}[c]{@{}l@{}}- 2 or more characteristics of the invention\\ - Include subject relative pronoun\\ - Include passive voice\end{tabular} \\ \hline
2024 Fall & Introducing hot spots in our area & \begin{tabular}[c]{@{}l@{}}- 2 or more reasons why I recommend this place\\ - Include objective relative pronoun\\ - Include anticipatory it + to-infinitive\\ - Include verb + object + to-infinitive\end{tabular} \\ \hline
\end{tabular}
\end{table*}

\section{Rubric}
\label{sec:rubric}

\begin{figure*}[t]
    \includegraphics[width=0.9\textwidth]{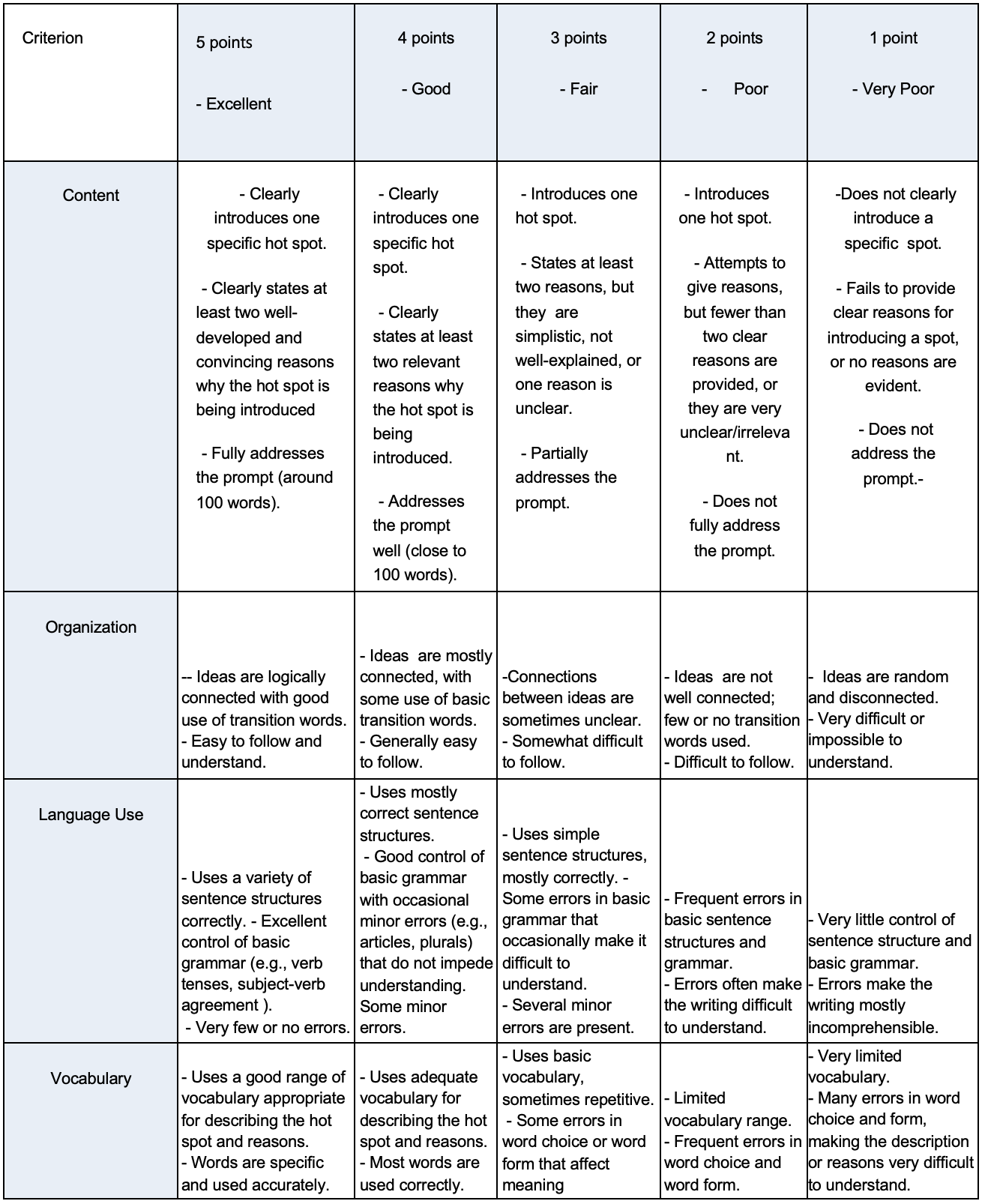}
    \caption{The rubric used to grade students' final assessment. The essays were graded using an analytic rubric adapted from \citet{jacobs1981testing}.}
    \Description{Figure showing the rubric used to grade students’ final assessment essays, adapted from Jacobs et al. Criteria include Content, Organization, and Grammar, each scored on a scale from 5 points (Excellent) to 1 point (Very Poor). Higher scores indicate a clear introduction of the topic, well-developed reasons, logical organization, and strong grammar with few errors. Lower scores indicate vague or missing reasons, disorganized ideas, frequent grammar errors, and incomprehensible writing.}

    \label{fig:rubric}
\end{figure*}

Figure \ref{fig:rubric} outlines the details of the rubric used for grading students' final assessments.

\section{Assessment Score}
\label{sec:score}

\begin{figure*}[t]
    \includegraphics[width=0.6\textwidth]{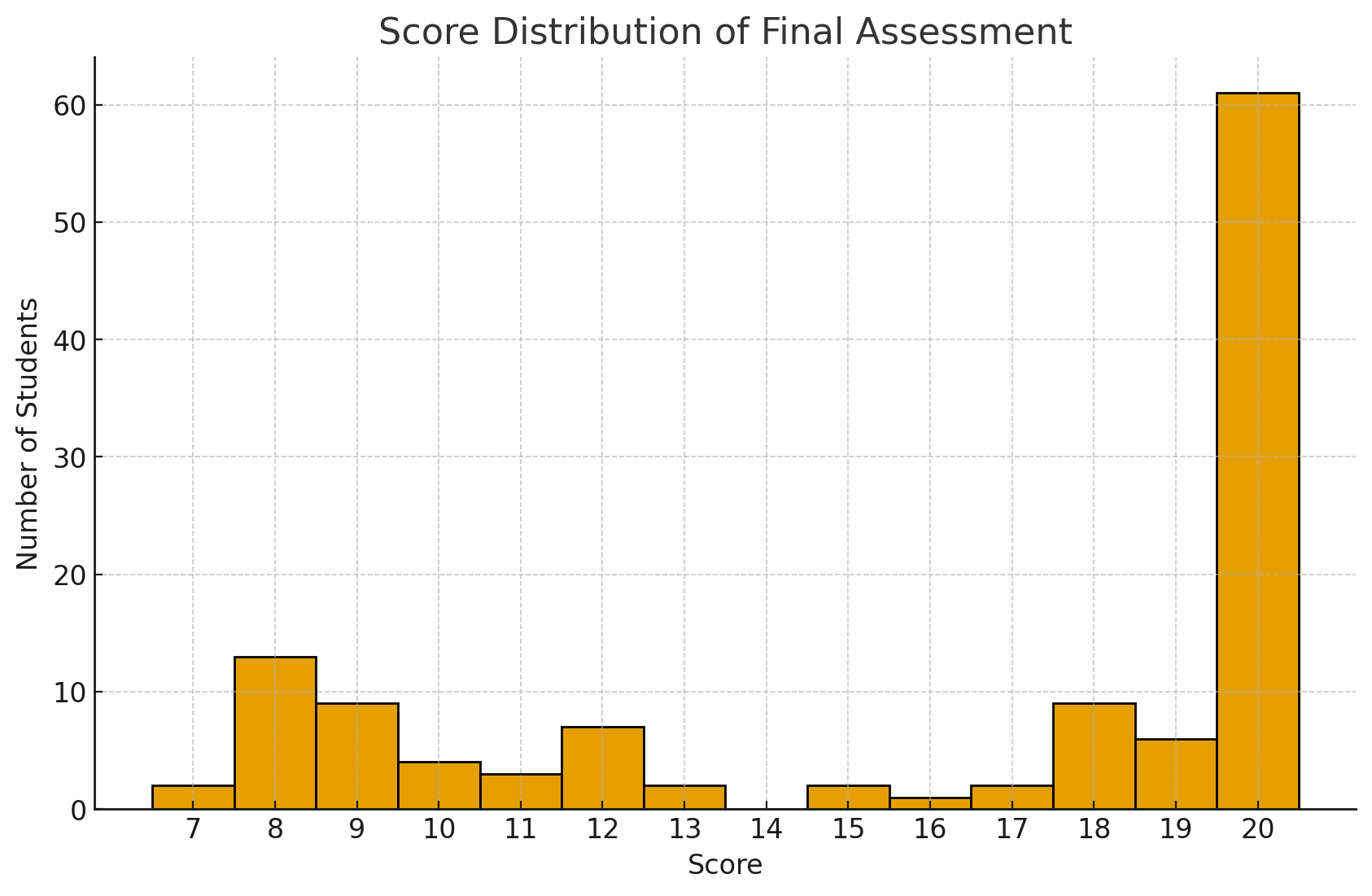}
    \caption{The score distribution of final assessment done in Fall 2024.}
    \Description{Histogram showing the score distribution of the final assessment in Fall 2024. The x-axis represents scores from 7 to 20, and the y-axis represents the number of students. Most students scored 20, with smaller groups clustered around 8, 12, 18, and 19.}
    \label{fig:score}
\end{figure*}

Figure \ref{fig:score} shows the distribution of the final assessment scores taken in Fall 2024.

\end{document}